\RequirePackage{fix-cm}
\documentclass[smallextended]{svjour3}       
\smartqed  
\usepackage{graphicx}
\usepackage{lscape}
\begin{document}

\title{The Data Zoo in Astro-WISE
}


\author{Gijs A. Verdoes Kleijn, Andrey N. Belikov, John P. McFarland}
%

\institute{
              Kapteyn Astronomical Institute, University of Groningen, P.O. Box 800, 9700 AV Groningen, The Netherlands
}


\date{Received: date / Accepted: date}

\maketitle

\begin{abstract}
In this paper we describe the way the Astro-WISE information system (or simply Astro-WISE) supports the data from a wide range of instruments and combines multiple 
surveys and their catalogues. Astro-WISE allows ingesting of data from any optical instrument, survey or catalogue, processing of this data to create 
new catalogues and bringing in data from different surveys into a single catalogue, keeping all dependencies back to the original data. Full data lineage 
is kept on each step of compiling a new catalogue with an ability to add a new data source recursively.
With these features, Astro-WISE allows not only combining and retrieving data from multiple surveys, but performing scientific data 
reduction and data mining down to the rawest data in the data processing chain within a single environment.
 
\end{abstract}

\section{Introduction}
\label{s:intro}

The increasing challenge of data management in astronomy is created not only by the increasing volume of the data flowing in from the telescopes, but as well by the variety 
of new astronomical catalogs and surveys being created. Combined analysis of multi-survey datasets from different instruments has the potential to 
solve outstanding questions in a wide range of astronomical and astrophysical areas. 
These include the combining of optical and near-infrared surveys to identify objects of interest such as very high-redshift 
Quasars (SDSS-UKIDSS, see~\cite{QSO} for example), unusual Brown Dwarfs~\cite{BD} and ultra-compact binaries~\cite{CVs}. 
The very same surveys can also be combined to constrain fundamental properties of our Universe: the nature of dark energy, 
the nature and distribution of dark matter via galaxy weak lensing, correlations in galaxy and QSO distributions and growth of large-scale structures. Complex relationships such as evolution of galaxies and their environment and nuclear 
activity require combined analysis from radio to X-rays (GAMA~\cite{GAMA}, AEGIS~\cite{AEGIS}, COSMOS~\cite{COSMOS}, 
Coma Legacy Survey~\cite{COMALS}, ATLAS3D~\cite{ATLAS3D}, GOODS~\cite{GOODS}, HUDF~\cite{HUDF}). 
The challenge for any information system hosting these datasets is no longer archiving of data products, but 
providing abilities to perform data mining and data reprocessing on a massive scale.  

The combining of a number of surveys and the data mining of the resulting combined survey is not a trivial task due to the volume of 
the data and to the particular requirements each user has for the combined catalogue. There are many ways to perform this task--from using the
abilities provided by the survey itself to employing resources and software of Virtual Observatory. 

The Virtual Observatory (VO) is the system of standards for publishing and accessing astronomical data developed 
by the International Virtual Observatory Alliance\footnote{http://www.ivoa.net/}. At the same time VO is the collection of all 
data available according to these standards. They are published by a number of organizations, for example, the European Virtual Observatory 
community\footnote{http://www.euro-vo.org/}. Recently VO standards were used not only to give data access but also to process 
the data itself by Canadian Advanced Network For Astronomical Research (CANFAR\footnote{http://canfar.phys.uvic.ca}).

Despite the progress achieved via the VO, combining of astronomical data from different surveys remains a challenge.
The problem with any of the methods is the limited volume of data which can be combined
into a new catalogue or archive and the detached nature of the produced
catalogue which is a separate entity without dependencies to the parent
surveys. In addition, the user must arrange for the storage of the produced data
and invent a way to automate the production of the new compiled catalogue.

These technical problems can been overcome, but not generally in the most efficient manner. Legacy research with large surveys often takes an ``\textit{end-point}'' approach: the end products of the surveys (e.g., calibrated images and catalogues) 
are taken in by the end-user as starting point for analysis. Combining just the catalogue end-products of surveys with $\leq 10^8$ entries is manageable 
using the infrastructure of a single user (``desktop science''). However, scientific requirements may lead 
to re-processing (e.g., going further backwards towards the raw data).
This can include homogenisation in terms of photometry, astrometry, 
aperture/image quality. As soon as the the analysis involves (re)processing of data and, hence, returning to the original images, 
it requires both hardware and software infrastructure well beyond the desktop regime to deal with the avalanche of 
heterogeneous/complex data. This reprocessing of the survey data often requires the full survey operation data flow system including the quality control task. In turn, this 
means that the user has to possess the similar human and computing resources as the original data processing site.

The natural way to solve these problems is by allowing the user to access the original data processing infrastructure. This gives 
the user an ability to reprocess data partially or fully to create his own version of the survey's end product. Nevertheless, this 
will not solve the problem of reprocessing another survey's data which the user would like to join with the first one. This requires a data processing system ideally allowing processing of data regardless of the survey/telescope/instrument/filters. Such a system 
must be designed to be generic enough so that the data from many different instruments can be handled by it. This system must have 
a generic pipeline with an ability to port new pipelines for new instruments and surveys.

The Astro-WISE information system~\cite{adass} is such system for optical/near-infrared wide-field imaging.
The Astro-WISE approach to the porting of new surveys and instruments allows the user to create a combination of surveys
and share it with other users. The key ingredient that allows the combination of different surveys and catalogues into new data products is 
a flexible common data model implemented in Astro-WISE. 
The ability to process the data inside Astro-WISE is defined by the level of integration of the external data in the Astro-WISE system. 
The deepest level of integration allows the processing and analysis of data from its most raw form, directly from the detector. At the shallowest level the scientific analysis starts in catalog-space using ingested external catalogues.

The difference between an astronomical data warehouse like the Virtual Observatory and Astro-WISE with its integration of wide range of surveys is the ability of (re-)processing and traceability of processing. The basis for this ability is a single, common data model specifically designed for this purpose. The integration takes place in an existing system that has a novel implementation of 
query language and tools for combining of catalogues. Astro-WISE provides the user with the necessary resources to do the job and 
allows the storage and sharing of the result of the compilation with both team members and the outside community.

In the next sections we describe the general design and components of the solution that allows Astro-WISE to be used as a platform to operate multiple 
surveys. The solution is based on the core principles of Astro-WISE: a common data model, data lineage and a modular approach to programming 
(see~\cite{WISE}).

\section{Common Data Model}

\begin{figure}
\includegraphics[width=0.75\textwidth]{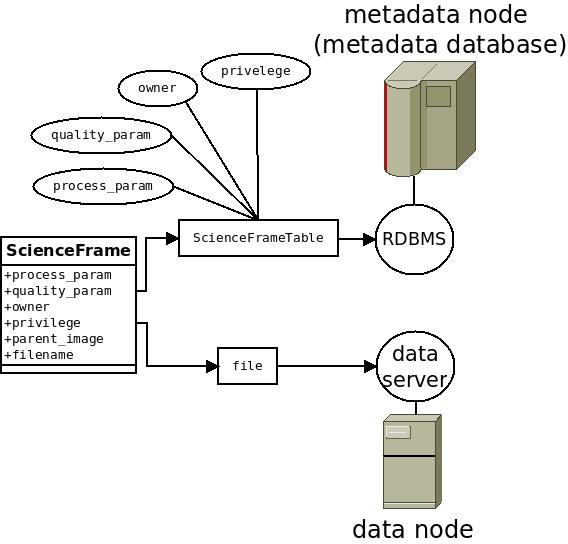}
\caption{
Data and metadata in the Astro-WISE approach. Data are stored as files on data nodes (data server), while metadata is stored in tables in metadata nodes (relational Database Management System). 
ScienceFrame can be any Astro-WISE class for images or calibration files (e.g., ReducedScienceFrame).  
}\label{fig:data_model}       
\end{figure}

\begin{figure}
\includegraphics[width=0.75\textwidth]{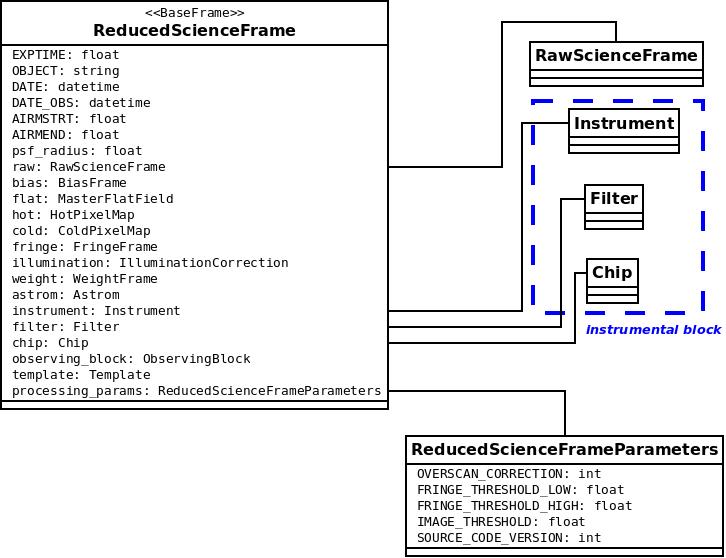}
\caption{
An example of the image frame in Astro-WISE data model. The ReducedScienceFrame stores links to all objects used to create 
the frame and separate the frame itself from instrumental block of Astro-WISE classes.  
}\label{fig:reduced_frame}       
\end{figure}

Before describing the solution to accommodate data of various format and nature, from raw images to catalogues spanning wavelengths 
from infrared to radio, we explain the technique used to store data in the Astro-WISE information system. Each data item in Astro-WISE 
can consist of two parts: pixel data and metadata. The pixel data is a multidimensional array which is stored in the FITS file and 
usually is an image or a calibration file. The metadata is a a full description of the pixel data, containing practically all 
information about the image apart from pixel array itself. This last requirement demands a lot from the data model which must accommodate all this information. For example, also catalogues are metadata and thus stored in the Astro-WISE database. Figure~\ref{fig:data_model} 
shows the basic principle of separation of the data and the metadata in Astro-WISE. This data model is realised on the metadata level, 
allowing modification by adding new attributes and classes without changing the existing data.

At the core of Astro-WISE is a \textit{data model} that is \textit{common} to all instruments acquiring a similar type of data, and is object-oriented. Initially, it was developed for the processing of optical 
data from only a few specific instruments (see~\cite{pipeline}). During the last decade, it evolved into a general and flexible tool to accommodate astronomical data from many instruments and catalog archives and can be adapted to others as needed. 
The imaging data--raw, reduced, regridded and coadded images--are described by corresponding Python classes separated from 
the classes describing instruments, filters, and detectors. The typical image object (an instance of a particular image class) has a link to the instrument, filter, and detector chip objects, but there are no 
instrument-specific attributes in the class, other than pixel size, describing the image itself. This feature allows the creation of an image object and all objects derived 
from the image object to be instrument-independent, thus allowing introduction of new instruments into the system without changing the data model itself. 
The same principle of separation of attributes is used for processing parameters.

Thus the instrument itself becomes conceptually a processing parameter in a single processing pipeline. 
This means that the user processes data from different surveys and instruments with the same software and interfaces. This pipeline 
integration is realised by using this \textit{common data model} for all the data stored and processed in the system.

Figure~\ref{fig:reduced_frame} illustrates the separation of attributes of the \texttt{ReducedScienceFrame} data item. The frame has 
a link to the ``parent'' frame (\texttt{RawScienceFrame}) along with links to all calibration frames used in the processing. The instrumental 
block consists of links to instances of classes \texttt{Instrument}, \texttt{Filter} and \texttt{Chip} which describe the features of the camera.

\section{Reusability of the pipeline} 

The modular approach to pipelining combined with the common data model makes the general optical image processing pipeline 
implemented in the Astro-WISE information system reusable for a vast number of instruments\footnote{See http://www.astro-wise.org/portal/instruments\_index.shtml for examples currently integrated into Astro-WISE}. 
The ability to use the same module to process the data from different instruments is due to the fact that all 
instrument-dependent parameters are grouped in special classes and instrument-dependent processing is done on the calibration level. 
For example, the overscan correction method used to normalize bias levels can be done in a number of different 
ways\footnote{See http://www.astro-wise.org/portal/howtos/man\_howto\_bias/man\_howto\_bias.shtml} 
and are specified by a single processing parameter.
This processing parameter is saved in the processing parameters of the image object to which it is applied. For example, the user specifies the target for the processing, a \texttt{ReducedScienceFrame} object, that should be produced from a \texttt{RawScienceFrame} object. The pipeline retrieves the instrument-specific 
parameters and calibration objects using instances of classes \texttt{Instrument}, \texttt{Filter} and \texttt{Chip}, referred to by the instance 
of the \texttt{RawScienceFrame} class.   

As a result, the integration of a new instrument into the Astro-WISE system is a simple creation of a new set of instances of \texttt{Instrument}, 
\texttt{Filter} and \texttt{Chip} classes and the creation of a \texttt{HeaderTranslator} class containing translations or mappings of all instrument-specific metadata stored with the pixel data.  This class homogenizes how the metadata enters the system so that all parameters have common designations regardless of the instrument (see Section~\ref{sec:ingest}).

The pipeline is reusable not only for new images ingested in the system but for new use-cases which involve the same data. The user 
can reprocess the data with new calibration images and processing parameters such as overscan correction method, detection threshold for source extraction,  etc.

\section{Access policy and visibility}

A major aim of Astro-WISE is to pool the data calibration efforts by its users. A calibration coverage that is continuous in time can be created collaboratively. This way, the calibration for a specific night of science data can be improved from trend analysis on calibration and science data over longer timespans. Astro-WISE is effectively calibrating the instrument, not individual nights or observations. This means that the calibration data should be shareable 
with other users in a flexible manner including the tracing to the original raw calibration data. Calibration objects (e.g., biases, flat fields, etc.) should become available for the lifetime of the instrument to the Astro-WISE community. Moreover, any user should be able to decide among different calibration sets which one is most applicable to his/her science data and science goals.

The access policy of Astro-WISE has in its core an ability to join users into project groups and share all the data in these projects between its members. Each user 
has his own space (termed MyDB) where all private processing is done, including calibrations. All the data in this MyDB space is accessible to the owner of the data only. To share the data within user's group, it must be \textit{published} to a more public level.  To facilitate this, each data item (object) has an attribute \texttt{\_privileges} that defines the scope of visibility of this data item. By default, the attribute 
is set to $privileges=1$ (MyDB) or to another preselected level. The user can manually raise the attribute to $privileges=2$ (publish to the project level: the data item is visible to all 
users within the project), $privileges=3$ (publish to the Astro-WISE level: the data item is visible to all Astro-WISE users), $privileges=4$ 
(publish to the world level: anonymous user can retrieve the data item with Astro-WISE interfaces) and, finally, $privileges=5$ 
(publish to the VO level: the data item is visible in the Virtual Observatory). 
The access policy allows users to share not only processed data but imported external data as well, products derived from it and/or associated with existing data.

\section{Data ingest}\label{sec:ingest}

\begin{figure}
\includegraphics[width=0.95\textwidth]{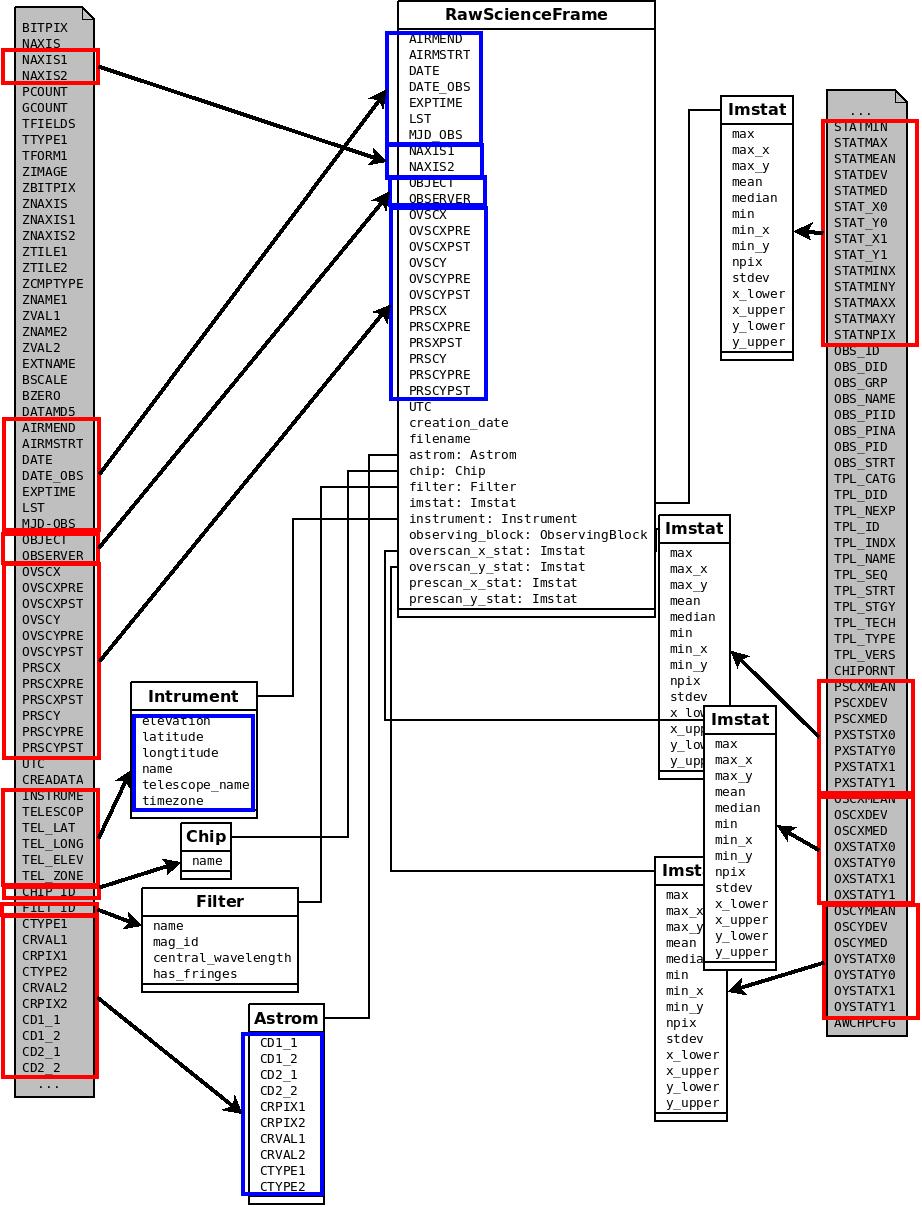}
\caption{
Mapping of the FITS file header with the content of a system of objects created for the raw data to be ingested. 
The \texttt{RawScienceFrame} has as attributes a number of references to objects which define astrometric parameters, image statistics, filter, instrument, etc.
Some of dependencies are omitted in the Figure.}
\label{fig:mapper}       
\end{figure}

Adding a new instrument and its data to the Astro-WISE information system starts with creating a \texttt{HeaderTranslator} (a mapper) that converts the set of data, primarily the metadata, in some native format (e.g., headers and 
pixel data in FITS format, pixelmaps, weight maps) coming from a wide-field imager into a form that maps onto the common data model. 
This ``ingestion'' process handles raw data in its native format directly from the instrument or archive and can be expanded to handle preprocessed data at any stage.  Ideally, the reliability should be such that an Astro-WISE user can blindly feed the mapper 
with data from any phase of an instruments' life. To give an idea what is involved in this 
mapping,  Fig.~\ref{fig:mapper} shows the mapping of the metadata content for a \texttt{RawScienceFrame}. 
It shows that the mapping involves setting approximately 30 of the \texttt{RawScienceFrame}'s properties and about 10 dependencies having 
tens of properties each. The translation populates a number of objects of different classes representing different aspects 
of the metadata associated with the \texttt{RawScienceFrame} object. The separate classes for instrument-specific and processing-specific parameters 
in the common data model eases the automatic population of the objects' attributes.  
If translation to the common data model becomes a manual operation it defeats the purpose of having a common model for common handling of 
wide-field imagers and their surveys. This means that the bookkeeping aspects of the 
translation should be robust against changes in telescope and instrument setup. 
The common data model has a single detector as 
its atomic unit. This facilitates handling of detector specific characteristics (e.g., usage of segmented filters) and changes therein (e.g., chip replacement). 
It also means that it should deal with idiosyncrasies of instruments ranging from changes in FITS 
header contents or naming conventions to erroneous header contents (e.g., bogus astrometry, photometry). 

The translator creates an environment where the data reduction and post-reduction data analysis applications operate uniformly across instruments. For data reduction, the quantitative definition of science and weight pixel data, and photometric and astrometric reference 
systems are all be brought to a single model representation in the common data model. This is a commonality that allows a single 
processing pipeline in which instrument effectively becomes a bookkeeping parameter, and in rare cases, a process configuration parameter at intake.

\section{Levels of Integration}

The common data model in the Astro-WISE information system spans a wide range data types from raw science and calibration observations to modelling results of sources 
in calibrated, stacked data. Mappers exist in Astro-WISE to map external data onto the common data model at various levels of end-user data products:
\begin{itemize}
\item{raw observations}
\item{de-trended data, photometrically, astrometrically calibrated stacked data}
\item{catalog data}
\end{itemize}
For the levels more processed than raw it is optional (not mandatory) to ingest also all data and metadata from the processing chain 
that led to it to incorporate that data lineage.

The need for the different levels of integration for external data is due to different use-cases for these data. 
The use-cases can be divided based on the level of reprocessing which the integration will allow. The most simple case is the ingestion of 
an external catalog ``as is''. This type of integration supposes no reprocessing of the data is needed. On the other end of the use-cases
is ingestion of raw science and calibration images allowing reprocessing of the data completely from scratch.

In more detail, the types of integration of external data are:
\begin{itemize}
\item[] {\bf raw data level} - both raw data with all necessary calibration files and the pipeline are imported into Astro-WISE 
allowing the user to reduce the data completely as new basis for e.g., source extraction; 
\item[] {\bf reduced data level} - only the processed images are ingested, the detrending and/or photometric and astrometric calibration of images is done (partially) outside Astro-WISE. This way, the user can extract sources from external images using his own preferences in extraction parameters; 
\item[] {\bf catalog level} - only the final catalog is ingested. The user can combine it with data of other surveys in the system.
\end{itemize}

Table~\ref{tab:1} shows the collection of surveys, catalogues and instruments that are integrated in Astro-WISE. As we can see, for a number of 
instruments Astro-WISE can be used as a data processing environment. This range of instruments is not fixed. The user of Astro-WISE can create a new mapper 
to ingest the data from an instrument not listed in the table and use the Astro-WISE pipeline to reduce the data.  
Accompanied by a modular approach to the processing pipeline, the integration of data from new instruments and surveys is feasible if it involves mapping onto the common data model as is or adding parameters to the common metadata description. 

\begin{table}
\caption{External Data currently in Astro-WISE}
\label{tab:1}       
\begin{tabular}{|c|c|c|}
\hline\noalign{\smallskip}
Level of integration & Instrument/Survey/Catalog \\
\noalign{\smallskip}\hline\noalign{\smallskip}
RAW & OMEGACAM@VST, WFI@ESO/MPG2.2m, WFC@INT \\
  & SUPRIMECAM@SUBARU,LBCBLUE@LBT, LBCRED@LBT,\\
    &  MEGACAM@CFHT, MDM8K@MDM, \\
    & GRATAMA@BLAAUW, MONICA@WENDELSTEIN \\
\hline
REDUCED & WSRT, ACS@HST, VIRCAM@VISTA, LOFAR \\
  & ISAAC@VLT, WFCAM@UKIRT, ESOSCHMIDTPLATES, PDS \\
\hline
SURVEY CATALOGS & USNO, SDSS, UKIDSS, 2MASS-PSC, ESO-LV \\
\noalign{\smallskip}\hline
\end{tabular}
\end{table}

The processing of data in Astro-WISE is based on the separation between pixel data and metadata. Pixel data in this sense are images\footnote{Pixel data includes raw images, 
calibration and calibrated images, and reduced images that come from the instrument as they are.} in FITS format 
ingested into the system without any changes. Metadata is the description of the pixel data stored in the header of FITS files. Catalogues are also metadata in Astro-WISE.
During processing, any information added to this initial metadata is mainly relationships between the raw, processed images, their processing parameters 
and references to calibration files. As long as the data processing can be mapped onto the 
same elementary blocks, as is generally the case for optical image processing, the same pipeline can be used for processing the data from different instruments.
Changes in detector/atmospheric behaviour or new insights do not change dependencies. Certain processing steps can be optional, such as
fringe correction for data taken in redder filters, global forms of astrometry and photometry, and illumination correction (photometric flat field).
And finally, despite a design specifically for optical data, it is possible to adapt a new pipeline to the optical data model, or one based on it, to accommodate data which requires special treatment,  
as would be the case for infrared data, for example.

\section{Examples of use-cases for the data integration}

\begin{figure}
\begin{center}
\includegraphics[width=\textwidth]{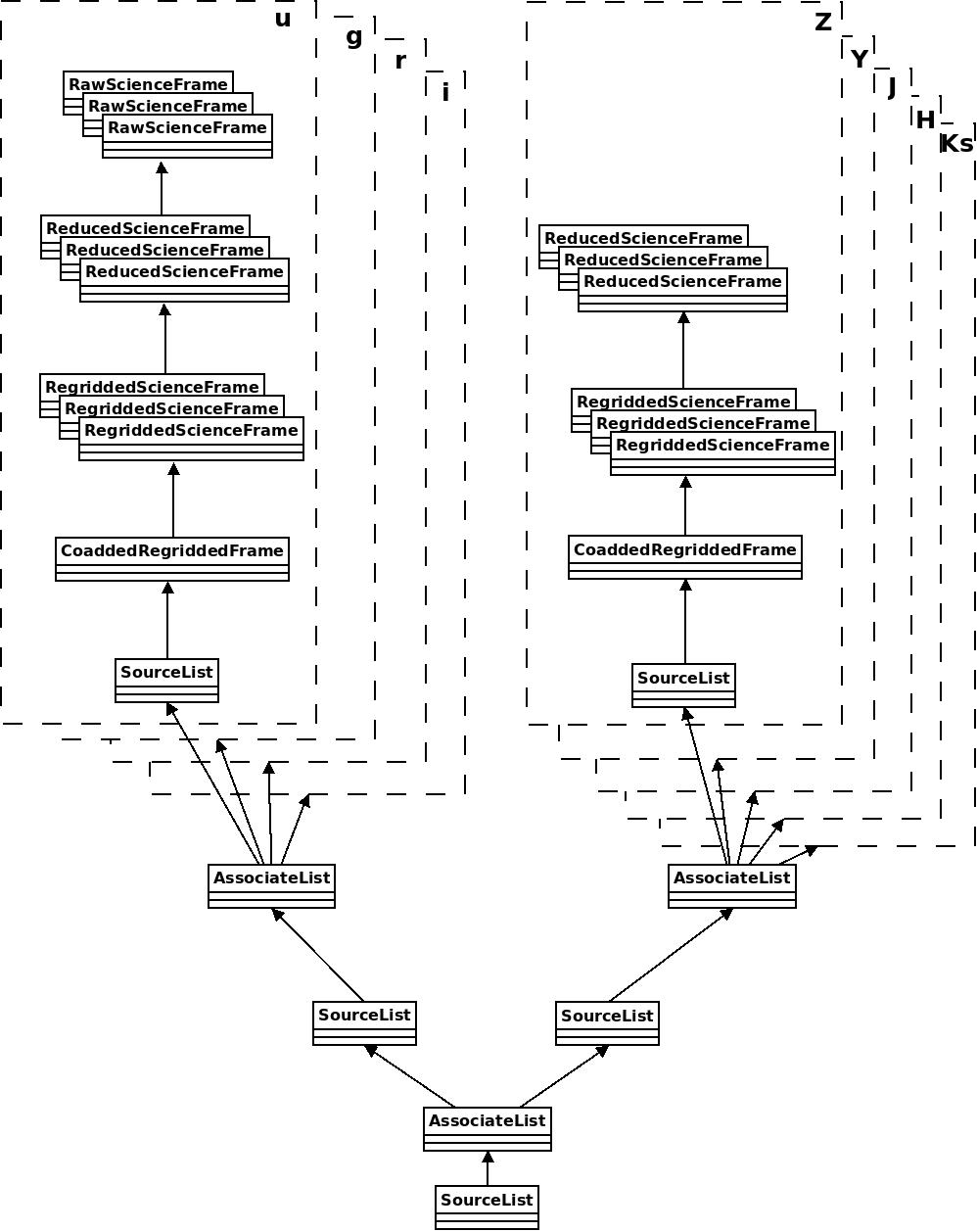}
\caption{
Chain of dependencies for the final multi-band catalog which joins observations for KiDS {\it ugri} bands and VIKING {\it ZYJH$\textrm{K}_S$} bands. In this case the VIKING 
data was ingested in Astro-WISE as reduced images (the product of CASU) and regridded to the KiDS grid points. 
}\label{fig:depend2}
\end{center}
\end{figure}
 
Apart from the obvious case of merging the data from multiple surveys into a dataset of effectively uniform quality, there are a number of use-cases 
which apply the techniques described above. We will show one of the simplest examples where the user can easily 
change a number of complicated relationships during data processing to profit from the common data model in the Astro-WISE information system.

\subsection{Merging of surveys: combining the KiDS and VIKING surveys}
\label{sec:merge}

The optical KiDS and near-infrared VIKING imaging surveys are twin surveys~\cite{KiDS}. They cover the same 1500 square degree area in different 
imaging bands. KiDS is using the OmegaCAM imager on the VLT Survey Telescope consisting of an array of 32 CCDs 
to observe the survey area with the $u'g'r'i'$ filter bands. VIKING is using the VIRCAM 
imager on the VISTA telescope, uses 16 detectors and covers the area with the $ZYJHK$ filter bands.  
KiDS raw-to-catalogue processing is led by OmegaCEN and is done using Astro-WISE. The VIKING survey is processed 
by CASU-Cambridge and WFAU-Edinburgh. The survey designs have been matched in terms of area and depth to address a range of astronomical scientific questions that rely on combined analysis of optical and infrared data. A primary goal is weak lensing tomography~\cite{Weak} using photometric redshifts for $10^8$ galaxies with an accuracy obtainable from the combined 9-band optical through near-IR SEDs. Photometric redshifts require very accurate relative photometry, and detailed knowledge of aperture and PSF for each band. For this reason, the detrended 
VIKING data is ingested into Astro-WISE. 

VIKING data can be ingested into Astro-WISE at three levels. First, the VIKING catalogues produced by WFAU are ingested and mapped onto the common data model. This way, 
VIKING source data is available to be combined, mined and visualized in combination with KiDS and other surveys (e.g., 2MASS, SDSS) using Astro-WISE's capabilities to combine, mine and 
visualize a catalogue.  Second, the VIKING tiles, photometrically and astrometrically calibrated stacked imaging, is mapped onto the common data model. This allows to application of homogeneous 
source extraction (e.g., matched aperture photometry, dual-image mode, etc.) and modelling (e.g., GALFIT and GALPHOT ) on galaxies in KiDS and VIKING data. Third, instrumentally detrended VIKING 
data is ingested into Astro-WISE. Only then can research requiring comparison of KiDS and VIKING data at the pixel level (e.g., galaxy colormaps) be performed. This 
ingestion level is also required for the phot-z pipeline for the combined KiDS+VIKING data.
The KiDS and VIKING data must be calibrated using a single photometric reference system and astrometric reference system and projection. Fig.~\ref{fig:depend2} shows dependencies between intermediate data products on the way to the final catalog starting at this ingestion level. For each source in the final catalog, it is possible to identify the pixels in the raw image in which this source was detected (see~\cite{subimage}). 

For surveys in the middle of survey production (like KiDS and VIKING), the combination of Astro-WISE's common data model and access to extreme data-lineage makes it possible to take advantage of data quality improvements immediately during survey operations~\cite{quality}. The pipelines~\cite{pipeline} mentioned above select their input via database queries. The common data model ensures that queries can be constructed for the latest or for earlier versions 
of survey products or for catalogues with specified source extraction configuration. This means the pipeline user has full control over which version of data 
from a survey is considered using the innate data lineage in Astro-WISE.

\subsection{Bringing surveys onto a common astrometric reference frame}
\label{sec:astreffrm}

\begin{figure}
\begin{center}
\includegraphics[width=0.95\textwidth]{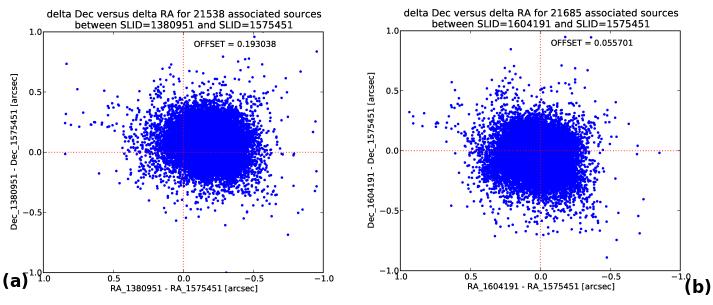}
\caption{
(a) Two-dimensional astrometric residuals between high-S/N ($S/N>10$) source extractions of a coadded MegaCam CARS field (in the W1 area) in $z'$-band using the USNO-B1.0 catalogue as the astrometric reference catalogue and high-S/N source extractions of a coadded VIRCAM VISTA field in $Z$-band using the Two-micron All-Sky Survey Point Source Catalogue (2MASS PSC) as the astrometric reference catalogue.  The two images cover nearly the same area and only source pairings from the common area are plotted.  It is clear from the locus of residuals that the two catalogues have a systematic offset of approximately 0.2 arcsec in this region of the sky. (b) Two-dimensional astrometric residuals between high-S/N ($S/N>10$) source extractions of a coadded MegaCam CARS field (in the W1 area) in $z'$-band using the Two-micron All-Sky Survey Point Source Catalogue (2MASS PSC) as the astrometric reference catalogue and high-S/N source extractions of a coadded VIRCAM VISTA field in $Z$-band using the 2MASS PSC as the astrometric reference catalogue.  The two images cover nearly the same area and only source pairings from the common area are plotted.  It is clear from the locus of residuals that the two observations are now on an equivalent astrometric reference frame.
}\label{fig:carscasu2d}
\end{center}
\end{figure}

During a search for photometric drop-outs and proper motion sources between the combined CARS MegaCam data in $ugriz$ and VISTA 
VIRCAM data in $ZYJHK$, significant astrometric discrepancies surfaced.
The origin of these discrepancies lies in the fact that the data from the CFHTLS W1 
field taken with the MegaCam instrument had been calibrated using the USNO-B1.0 
astrometric reference catalogue, while similar observations with the VIRCAM instrument were calibrated using the Two-micron All-Sky Survey Point Source Catalogue (2MASS PSC).  
A systematic offset of approximately 0.2 arcsec was noted for the area compared.  Recalibrating the MegaCam data to the same 
reference catalogue after coaddition brought the systematic offset to 
less then 0.06 arcsec, well below the formal errors of reference catalog.  
Figure~\ref{fig:carscasu2d} shows the results of the recalibration. 


\section{Conclusion and Future Work}

The Astro-WISE information system allows data handling for multi-survey operations and research within a single environment. The handling spans the from calibration of raw survey data to a wide range of post-calibration analysis (morphometry, photometric redshifts, variability etcetera). The special feature of the Astro-WISE information system is the ability to keep all processing dependencies inside the system. No matter how complicated the data 
lineage is, the user can understand how the final catalogue is compiled via the sequence of operations starting from the raw data, and can reprocess it using different processing parameters. Given that Astro-WISE keeps the quality parameters for each data item in the processing chain, it becomes possible to provide the user with everything necessary to accomplish multi-survey quality control and research inside a single system. 

The system is scalable in terms of imaging data sources (and hence in data volume). The counter currently stands at 18 different instruments which are supported in Astro-WISE. The user can add new instruments and surveys, and reuse code to reduce the new data sets and combine them with existing survey products. The key method that allows this achievement is the description of wide-field imaging data and instruments using a common data model. Astro-WISE allows (re)processing and analysis of an unprecedented wide range of surveys and archives with all operations done on the multiple surveys in a uniform manner.

Another advantage of the common data model is that the same programs and interfaces can be reused on new data, reducing time the user should spent on an adaptation to the new instrument or survey. 
This also saves time for interface building for new data sets~\cite{interface}.  

Presently Astro-WISE accommodates all types of astronomical catalogues varying from radio to X-rays. Source extraction and subsequent modelling and analysis can be done on radio, 
near-infrared and optical imaging. Optical and near-infrared imaging data at raw or any process level can be (re)processed within the system. Next steps in the development of the common 
data model are adaptation to accomodate spectroscopic data to make it possible to combine optical and spectroscopic surveys. The Astro-WISE approach for information processing has proven 
to be generic enough to be applied for radio survey handling (LOFAR~\cite{lofar}), optical multi-unit spectroscopy (Multi-Unit Spectroscopic Explorer) and beyond astronomy in the fields of 
medicine and artificial intelligence\footnote{http://www.rug.nl/target}.

\begin{acknowledgement}
The authors are very pleased to acknowledge the pioneering work by Edwin A. Valentijn, Kor G. Begeman, Danny R. Boxhoorn, Erik R. Deul and Roeland Rengelink in creating the OmegaCAM data model. This work laid the basis for the common data model in Astro-WISE.
\end{acknowledgement}



\begin{thebibliography}{}
%
%
\bibitem{adass}
    Valentijn, E.A., McFarland, J.P., Snigula, J., Begeman, K.G., Boxhoorn, D.R., Rengelink, R., Helmich, E., Heraudeau, P., Kleijn, G.V., Vermeij, R., Vriend, W.-J., Tempelaar, M.J., Deul, E., Kuijken, K., Capaccioli, M., Silvotti, R., Bender, R., Neeser, M., Saglia, R., Bertin, E., Mellier, Y.: Astro-WISE: Chaining to the Universe. ASP Conference Series, Vol. 376, p.491 (2007)
\bibitem{ATLAS3D} 
Cappellari, M., Emsellem, E., Krajnovi{\'c}, D., et al.\ ,The ATLAS$^{3D}$ project - I. A volume-limited sample of 260 nearby early-type galaxies: science goals and selection criteria, MNRAS, 413, 813 (2011)
\bibitem{COMALS} 
Carter, D., Goudfrooij, P., Mobasher, B., et al.\ , The Hubble Space Telescope Advanced Camera for Surveys Coma Cluster Survey. I. Survey Objectives and Design, ApJS, 176, 424 (2008)
\bibitem{AEGIS} 
Davis, M., Guhathakurta, P., Konidaris, N.~P., et al.\ , The All-Wavelength Extended Groth Strip International Survey (AEGIS) Data Sets, ApJL, 660, L1 (2007)
\bibitem{GOODS} 
Giavalisco, M., Ferguson, H.~C., Koekemoer, A.~M., et al.\ ,The Great Observatories Origins Deep Survey: Initial Results from Optical and Near-Infrared Imaging, ApJL, 600, L93 (2004)
\bibitem{QSO}
Mortlock D.J., et al., A luminous quasar at a redshift of $z = 7.085$, Nature 474, 616 (2011)
\bibitem{BD}
Berger E. et al., Discovery of radio emission from the brown dwarf LP944-20, Nature 410, 338 (2001) 
\bibitem{CVs}
G\"ansicke B.T., SDSS unveils a population of intrinsically faint cataclysmic variables at the minimum orbital period, MNRAS 397, 2170 (2009)
\bibitem{WISE}
Begeman K., Belikov A.N., Boxhoorn D., Valentijn E., The Astro-WISE datacentric information system, Experimental Astronomy, submitted (2012) 
\bibitem{GAMA}
Hill, D.~T., Kelvin, L.~S., Driver, S.~P., et al.\ , Galaxy and Mass Assembly: FUV, NUV, ugrizYJHK Petrosian, Kron and S{\'e}rsic photometry, MNRAS, 412, 765 (2011) 
\bibitem{COSMOS} 
Scoville, N., Aussel, H., Brusa, M., et al., The Cosmic Evolution Survey (COSMOS): Overview, ApJS, 172, 1 (2007)
\bibitem{HUDF} 
Yan, H., Dickinson, M., Stern, D., et al.,Rest-Frame Ultraviolet-to-Optical Properties of Galaxies at $z \sim 6$ and $z \sim 5$ in the Hubble Ultra Deep Field: From Hubble to Spitzer, ApJ, 634, 109 (2005)
\bibitem{KiDS}
de Jong J. et al., Kilo Degree Survey, Experimental Astronomy, in press
\bibitem{pipeline}
McFarland J.P., Verdoes-Kleijn G., Sikkema G., Helmich E.M., Boxhoorn D.R., Valentijn E.A., The Astro-WISE Optical Image Pipeline: Development and Implementation, 
Experimental Astronomy, in press, arXiv:1110.2509v2
\bibitem{quality}
McFarland J.P., Helmich E.M., Valentijn E.A., The Astro-WISE Approach to Quality Control for Astronomical Data, Experimental Astronomy, in press
\bibitem{subimage}
Mwebaze J., Boxhoorn D., McFarland J., Valentijn E.A., Sub-image data processing in Astro-WISE, Experimental Astronomy, in press
\bibitem{Weak}
Mellier Y., Probing the Universe with Weak Lensing, ARA\&A, 37, 127 (1999) 
\bibitem{interface}
Belikov A.N., Vriend W.-J., Sikkema G., Astro-WISE interfaces: Scientific information system brought to the user, Experimental Astronomy, in press
\bibitem{lofar}
Begeman K., Belikov A.N., Boxhoorn D.R., Dijkstra F., Holties H., Meyer-Zhao Z., Renting G.A., Valentijn E.A., Vriend W.-J., LOFAR Information System, Future Generation Computing Systems 27, 319 (2011) 
\end{thebibliography}
\end{document}